\documentclass[debug,overfull,info]{epl}%
\usepackage{graphicx}
\usepackage{amsmath}
\usepackage{amsfonts}
\usepackage{amssymb}%
\shorttitle{Viscoelastic fractures in stratified composite materials}
\institute{
Physique de la Mati\`{e}re Condens\'{e}e, Coll\`{e}ge de France,
11 place Marcelin-Berthelot, 75231 Paris cedex 05, France\\
Department of Physics, Graduate School of Humanities and Sciences,
Ochanomizu University, 2--1--1, Otsuka, Bunkyo-ku, Tokyo 112-8610, Japan
}
\pacs{46.50.+a}{Fracture mechanics, fatigue and cracks}
\pacs{61.30.Dk}{Continuum models and theories of liquid crystal structure}
\pacs{83.10.Ff}{Continuum mechanics in rheology}
\begin{document}

\title{Viscoelastic fractures in stratified composite materials: "lenticular trumpet"}
\author{Ko Okumura\thanks{E-mail: okumura@phys.ocha.ac.jp}}
\maketitle

\begin{abstract}
We consider fractures in a stratified composite material with solid layers
separated by thin slices of extremely soft matter. Viscoelastic effects
associated with the soft layers are taken into account via the simplest model
for weakly cross-linked polymers. We find that certain small cracks running
along layers take a new "trumpet" shape quite different from previously known shapes.

\end{abstract}

\section{Introduction}

\begin{figure}[ptbh]
\begin{center}
\includegraphics[scale=0.5]{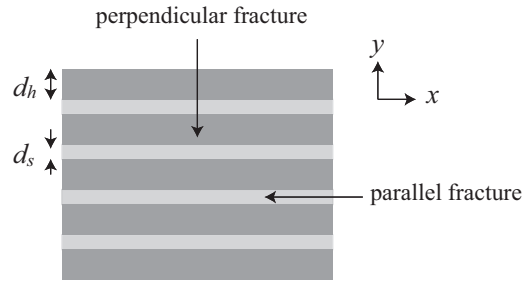}
\end{center}
\caption{Narce-type structure of materials: hard layers (elastic modulus
$E_{h}$, thickness $d_{h}$) are glued together by soft layers (modulus $E_{s}%
$, thickness $d_{s}$). Cracks in the $y-z$ plane and the $x-z$ plane are
called a perpendicular and parallel fractures, respectively (the $z$-axis is
perpendicular to this page).}%
\label{f1}%
\end{figure}

Nacre is a composite material with solid layers (aragonite) separated by thin
slices of soft organic matter (proteins) (Fig. \ref{f1}). It has a spectacular
toughness where a fracture propagates normal to the layers
\cite{nacre1,nacre2,nacre3,nacre4}. This toughness can be explained from the
absence of stress concentration in this structure \cite{v1,v2}. In the present
note, we consider a different problem, where the fracture plane is parallel to
the layers. The fracture properties have some similarity with those of a
smectic liquid crystal, which have been discussed under the name of lenticular
fracture \cite{lenticular,v3}. Our aim here is to consider the dynamics, when
the soft layers behave like a viscoelastic fluid. But we shall start with a
reminder of the statics, using a simple scaling argument.

The crucial feature is that, for a fracture cavity of size $X$ along the
layers, the elastically distorted zone has a size $Y$ (perpendicular to the
layers) which does not scale like $X$, but rather like%
\begin{equation}
Y\simeq X^{2}/l \label{y-x2}%
\end{equation}
\textit{for small cracks} ($X\ll l$), where $l$ is a characteristic length%
\begin{equation}
l^{2}=K_{B}/E_{0}.
\end{equation}
Here $K_{B}$ is a bending modulus and $E_{0}\left(  =\varepsilon E_{s}\right)
$ is an elastic modulus associated with soft layers. It is emphasized that
\textit{Eq. (\ref{y-x2}) always holds at the scaling level}. (For larger
cracks ($X>l$) the relation (\ref{y-x2}) turns back to $X\simeq Y$; this
regime will be discussed elsewhere \cite{full}).

For stratified materials, the following relation has been shown \cite{v3}:%
\begin{equation}
l=d/\sqrt{\varepsilon}%
\end{equation}
where%
\begin{equation}
\varepsilon=\frac{E_{s}}{E_{h}}\cdot\frac{d_{h}}{d_{s}}%
\end{equation}
in the small $\varepsilon$ limit (see Fig. \ref{f1} for notations). In the
case of nacre, the small crack condition $X\ll l$ is rather severe
($l\simeq50d$). Thus, the following considerations might be more practical for
some artificially synthesized layered composites where the soft part is a
\textit{weak gel}, $E_{0}$ is very small and $l\gg d$.

The small crack condition $X\ll l$ makes the situation different from smectic
liquid crystals where we have $Y\gg X$ because $l$ corresponds to an atomic
scale ($X/l\gg1$ in Eq. (\ref{y-x2})) \cite{lenticular}. In the present case,
for $X/l\ll1$, we have $X\gg Y$; \textit{the anisotropic strain field is
distributed widely in the }$x$\textit{ direction compared with in the }%
$y$\textit{ direction} (see Fig. \ref{f2} below).

The potential energy (per unit length in the $z$-direction) of the crack is%
\begin{equation}
F\simeq K_{B}\left(  \frac{u}{X^{2}}\right)  ^{2}XY-\sigma uX+GX
\end{equation}
Here, $u$ is the displacement in the $y$-direction, $\sigma$ the pulling
stress (along the $y$ axis) and $G$ the fracture energy (energy required to
create a new unit area). (It has been shown that the dominant component in
strain and stress tensors are indeed the $y$ components ($u$ and $\sigma$,
respectively) \cite{v3}). Note here the first elastic term can be equally
expressed as $E_{0}(u/Y)^{2}$ due to Eq. (\ref{y-x2}); this term actually
results from a local balance between these two elastic contributions and Eq.
(\ref{y-x2}) originates from this balance condition \cite{lenticular}.
Minimizing $F$ with respect to $u,$ then, $F(X)$ has a maximum defining the
onset of fracture. At this critical of fracture we have%
\begin{equation}
u\sim X,\;\sigma\sim X^{-1},\;\sigma u\simeq G \label{u-s}%
\end{equation}
where the last equation announces that \textit{the product }$\sigma u$\textit{
gives the fracture energy}. These fractures are very different from
conventional parabolic fractures in linear elastic fracture mechanics
\cite{LEFM}: $u\sim X^{1/2}$ and $\sigma\sim X^{-1/2}$. In a more precise
analysis, we have obtained forms for $u(x,y)$ and $\sigma(x,y)$, \cite{v3}
which are consistent with Eqs. (\ref{u-s}) and (\ref{y-x2}).

Eq. (\ref{y-x2}) states that, \textit{at the scaling level, a point away from
the tip by a distance }$X$\textit{ along the }$x$\textit{ axis has the same
order of strain (or stress) with that by a distance }$Y$\textit{ along the
}$y$\textit{ axis, when }$Y\simeq X^{2}/l$; the situation can be
\textit{conceptually represented} as in Fig. \ref{f2}.\begin{figure}[ptbh]
\begin{center}
\includegraphics[scale=1.1]{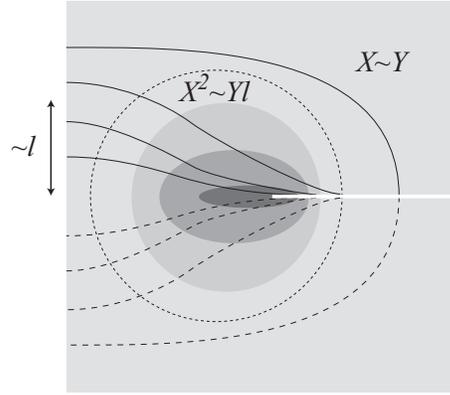}
\end{center}
\caption{Conceptual (rough) image of strain (stress) and deformation fields.
Solid line and broken lines are positive and negative deformation contours.
Magnitudes of stress or strain are indicated by a gray scale (the darker area,
the larger stress). Smaller the distance from the tip, the more anisotropic.
Beyond a distance larger than $l$, recovers the usual isotropic behavior.
(Fracture studied in this paper is limited: $\ X\ll l$). }%
\label{f2}%
\end{figure}

We now proceed to the dynamics via a complex modulus of the form
\cite{lenticular}%
\begin{equation}
\mu(\omega)=E_{0}+(E_{\infty}-E_{0})\frac{i\omega\tau}{1+i\omega\tau}
\label{visc}%
\end{equation}
where the ratio $E_{\infty}/E_{0}=\lambda$ is assumed to be large as it is in
a weakly cross-linked system. Here, $E_{0}$ is related to a small modulus
associated with weak cross-links while $E_{\infty}$ to a large modulus
originating from entanglements. Eq. (\ref{visc}) is a result of taking
viscoelastic effects in soft layers through a complex modulus of the same form
(but with $E_{0}$ and $E_{\infty}$ replaced by $E_{s}$ and $\lambda E_{s}$)
\cite{full}.

\section{Crack shape and fracture energy}

We consider a parallel crack propagating with a constant speed $V$. When $V$
is smaller than a sound velocity ($\simeq\sqrt{E_{0}/\rho}$), the equation of
motion of the density $\rho$ and the local velocity $v$ at the scaling level,
$\rho Dv/Dt=-\nabla\sigma$, reduces to a static equation: $\nabla\sigma=0$.
This is even true in our linear rheological model. In addition, the stress
components must also satisfy a compatibility equations, which directly result
from the definitions of strain fields as derivatives of deformation fields.
But these geometric conditions, again, have the same scaling structures for
our linear rheological model. Thus, \textit{the scaling relation for the
stress in Eq. (\ref{u-s}), }$\sigma\sim1/X\sim1/\sqrt{Y}$\textit{, remains
unchanged even in our viscoelastic model}. These observations will be more
precisely addressed elsewhere \cite{full}.

Another important observation here is the scaling identification of a distance
$X$ \textit{along the }$x$\textit{ axis} from the tip and the frequency
$\omega$ via the speed $V$:
\begin{equation}
X\simeq V/\omega; \label{rV}%
\end{equation}
small distances correspond to high frequencies while long distances to low
frequencies --- the farther away from the tip, the more time for relaxation.
In addition, from Eq. (\ref{y-x2}), the same magnitude of strain with the
point specified by the above $X$ is developed at a point separated from the
tip by a distance $Y\left(  \simeq X^{2}/l\right)  $ along the $y$ axis; these
points have the same time ($1/\omega\simeq V/X$) to relax. Thus, a distance
$Y$ from the tip sees the same frequency frequency $\omega\ $but a slower
propagating speed $V_{y}\simeq VX/l$ ($Y\simeq V_{y}/\omega$). In other words,
we can imagine that Fig. \ref{f2} correspond to a sequential propagation (from
the center to outwards). We also note that, seen from a new coordinate
$(x^{\prime},y^{\prime})=(x^{2}/l,y)$, the system returns to an isotropic
system. For example, Eqs. (\ref{u-s}) are changed into the conventional
parabolic form: $\sigma\sim1/\sqrt{X^{\prime}}\sim1/\sqrt{Y^{\prime}}$ and
$u\sim\sqrt{X^{\prime}}\sim\sqrt{Y^{\prime}}$ etc.

\begin{figure}[ptbh]
\begin{center}
\includegraphics[scale=1.9]{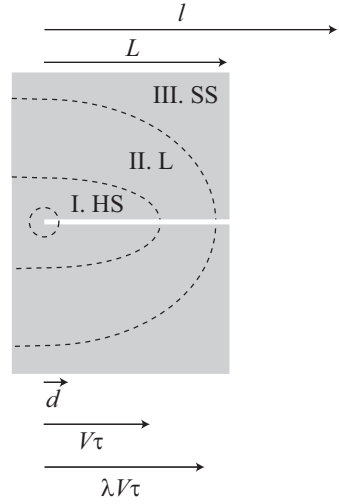}
\end{center}
\caption{Three spatial regimes for our viscoelastic model. Due to the scaling
relation $X^{2}\simeq Yl$, the regions appear as anisotropic. The smallest
region defined by $d$ comes out for a continuum theory. Note that for a crack
smaller than in this figure ($V\tau<L<\lambda V\tau$) only the regions I and
II are developed while for an even smaller crack ($d<L<V\tau$) only the region
I is developed; this figure corresponds to a fully developed crack ($L>\lambda
V\tau$). Note also that the degree of anisotropy of the boundaries separating
regions are subject to the ratio of $V\tau$ or $\lambda V\tau$ to the length
$l$; the smaller the ratio, the more anisotropic.}%
\label{f3}%
\end{figure}

Our viscoelastic model in Eq. (\ref{visc}) has three regimes depending on
frequencies: (I) at small frequencies ($\omega\tau\ll1/\lambda$), it is like a
soft solid with a small modulus $\mu(\omega)\simeq E_{0}$, (II) at
intermediate frequencies ($1/\lambda\ll\omega\tau\ll1$), it is like a liquid
with viscosity $\mu(\omega)\simeq i\omega\eta$, and (III) at high frequencies
($1\ll\omega\tau$), it is like a solid with a large modulus $\mu(\omega)\simeq
E_{\infty}$. Due to the correspondence between a distance and a frequency in
Eq. (\ref{rV}), a fracture can be thus spatially divided into three regions
(Fig. \ref{f3}):%
\begin{equation}%
\begin{array}
[c]{ll}%
\text{(I) }\lambda V\tau\ll X\text{:} & \text{soft solid (modulus }%
E_{0}\text{)}\\
\text{(II) }V\tau\ll X\ll\lambda V\tau\text{:} & \text{liquid (viscosity }%
\eta\text{)}\\
\text{(III) }d\ll X\ll V\tau\text{:} & \text{hard solid (modulus }E_{\infty
}\text{)}%
\end{array}
\text{ }%
\end{equation}

We consider a small crack size $L$ with $L\ll\lambda V\tau$ $\ll l$ (larger
crack sizes ($L\gg l$) will be discussed elsewhere \cite{full}); then, all the
three regions (I)-(III) are fully developed. The soft-solid region (I)
corresponds to low frequencies, and thus, to the static limit; in this region
($l\gg X\gg\lambda V\tau$), Eqs. (\ref{u-s}) hold. In the liquid zone (II),
the stress field scales as $\sigma\simeq\omega\eta u/Y\simeq\eta Vlu/X^{3}$
and at the same time it should scale as $\sigma\sim Y^{-1/2}\sim1/X$ even in
this liquid region as stated above. Thus, the strain should scale as $u\sim
X^{2}$ and the product as $\sigma u\simeq X$. The coefficients can be
determined by the matching at $X\simeq\lambda V\tau$ (the boundary between I
and II):
\begin{equation}
\sigma u\simeq\frac{GX}{\lambda V\tau}\text{ \ (liquid zone)}. \label{suL}%
\end{equation}
In the hard-solid region of $E_{\infty}$, we find $\sigma\simeq1/\sqrt{Y}$ and
$u\simeq\sqrt{Y}$ as in Eqs. (\ref{u-s}) but with $\sigma u\simeq G_{0}$ (via
the same manner as in deriving Eqs. (\ref{u-s})). Here, $G_{0}$ is associated
with the hard solid appearing near the tip. Matching this latter product
$\sigma u$ with that in Eq. (\ref{suL}) at $X\simeq V\tau$, we find
$G\sim\lambda G_{0}$. \textit{The overall separation energy }$G$\textit{ for a
fully developed crack is enhanced from }$G_{0}$\textit{ associated with local
precesses near the tip}. Note here that this expression for $G$ is valid for
$d<V\tau$ in our continuum theory.

\begin{figure}[ptbh]
\begin{center}
\includegraphics[scale=0.9]{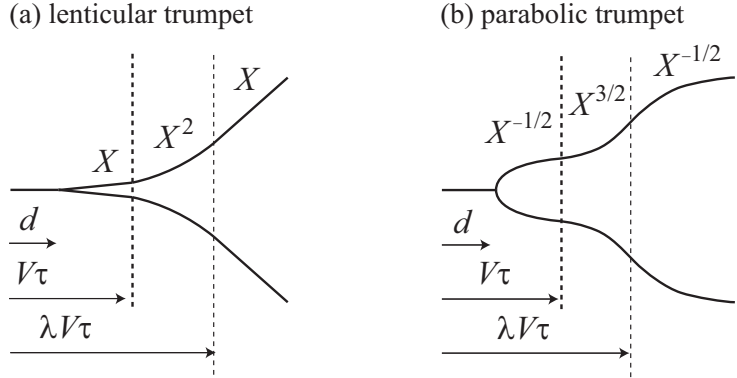}
\end{center}
\caption{Lenticular and parabolic viscoelastic trumpets.}%
\label{f4}%
\end{figure}

The crack shape resulting from this analysis is summarized as follows:%
\begin{equation}
u\sim\left\{
\begin{array}
[c]{ll}%
X & \text{ for }\lambda V\tau<X\\
X^{2} & \text{ for }V\tau<X<\lambda V\tau\\
X & \text{ for }X<V\tau
\end{array}
\right.
\end{equation}
It is just like a trumpet with a lenticular edge (Fig. \ref{f4}), as has been
suggested by the name of the model, but different from previously known
shapes; it is not similar to the conventional parabolic form nor an isotropic
parabolic trumpet predicted \cite{trumpet} and observed \cite{Ondarcuhu} in
certain polymer systems (Fig. \ref{f4}).

We complete our arguments by considering smaller fractures. When
$V\tau<L<\lambda V\tau$, only the hard-solid and liquid region are present;
the soft solid has yet to develop. In this situation, the fracture energy is
given by Eq. (\ref{suL}) at $X=L$: $G_{0}L/\left(  V\tau\right)  $; the
toughness decreases with velocity. When $L<V\tau$, only the hard-solid region
is developed and the fracture energy is given by $G_{0}$. Thus, with increase
in $V$, the fracture energy starts from a larger plateau value $\lambda G_{0}%
$, and then decreases to reach a smaller plateau value $G_{0}$:
\begin{equation}
G(V)\simeq\left\{
\begin{array}
[c]{ll}%
\lambda G_{0} & \text{ for }d/\tau<V<L/(\lambda\tau)\\
G_{0}L/\left(  V\tau\right)  & \text{ for }L/(\lambda\tau)<V<L/\tau\\
G_{0} & \text{ for }L/\tau<V
\end{array}
\right.  \text{ }%
\end{equation}

This behavior can be confirmed more precisely from a general formula:%
\begin{equation}
\frac{G(V)}{G_{0}}\simeq E_{\infty}\int\frac{d\omega}{\omega}\operatorname{Im}%
\left[  \frac{1}{\mu(\omega)}\right]  \label{GF}%
\end{equation}
which can be analytically calculated for the present model (just as in the
previously known isotropic case \cite{Florent}). It should be emphasized here
that this formula is unaltered even in our anisotropic materials, which is by
no means trivial; Eq. (\ref{GF}) here can be shown in the following manner. We
start from a relation:%
\begin{equation}
G(V)V\simeq\int dx\int dy\sigma\dot{e}\simeq\int dXY\sigma_{X}\dot{e}_{X}.
\end{equation}
In order to estimate $\dot{e}_{X}$ we again use the dimensional identification
in Eq. (\ref{rV}): $\dot{e}_{X}\simeq\dot{e}_{\omega}\simeq\omega
\sigma_{\omega}/\mu(\omega)\simeq\omega\sigma_{X}/\mu(\omega)$ and
\begin{equation}
G(V)\simeq\int d\omega\frac{Y\sigma_{X}^{2}}{\omega\mu(\omega)}\simeq
E_{\infty}G_{0}\int d\omega\frac{1}{\omega\mu(\omega)}%
\end{equation}
Here, we have used a more precise form of Eq. (\ref{u-s}): $\sigma_{X}%
\simeq\sqrt{E_{\infty}G_{0}/Y}$ \cite{full}. Since the real and imaginary part
of $1/\mu(\omega)$ are even and odd functions, respectively, we arrive at Eq.
(\ref{GF}).

\section{Conclusion}

In this paper, we present a physical picture for fractures in nacre-type
materials via scaling arguments. Viscoelastic effects for parallel fractures
are taken into account via the simplest viscoelastic model for weakly
cross-linked polymer. We expect that for slow crack-propagation speeds
($l\gg\lambda V\tau$) a small crack ($L\ll l$) takes a new trumpet shape
different from previously reported shapes; in the opposite limit will be
discussed in a separate paper \cite{full}. The overall fracture energy $G$ is
found to decrease from a larger plateau value $\lambda G_{0}$ to a smaller
plateau value $G_{0}$ with increase in velocity $V$ where $G_{0}$ is
associated with local precesses near the tip.

\begin{acknowledgments}
K.O. is grateful to P.-G. de Gennes for fruitful discussions and also for
reading the drafts prior to submission with giving useful comments.
K.O. also appreciates Elie Rapha\.{e}l and Florent Saulnier for discussions.
K.O. thanks members of the group of P.G.G. at
Coll\`{e}ge de France, including David Qu\'{e}r\'{e}, for a warm hospitality
during his third stay in Paris. This stay is financially supported by
Coll\`{e}ge de France.
\end{acknowledgments}

\end{document}